# Reversibilities and irreversibilities in thermoelectric energy conversion

Tom Markvart
Centre for Advanced Photovoltaics,
Faculty of Electrical Engineering,
Czech Technical University,
166 36 Prague 6, Czech Republic

**Abstract:** Similarly to Thomson, we consider the thermoelectric generator at open circuit as a classical heat engine. It is shown that, as long as the Thomson coefficient is nonzero, the operation generates entropy and is therefore irreversible. By expanding Thomson's approach we show that the voltage produced can be described by the usual Guy-Stodola equation for classical heat engines.

Debates about reversibility have been at the heart of thermoelectricity since the first steps towards theoretical elucidation.[1] Thomson (later lord Kelvin) obtained the fundamental theoretical relations using classical thermodynamics by arguing that a carrier of electricity (electron) also carries heat that can be converted into work.[2] Following several decades of discussions (see, for example, Refs. 3 and 4), a consensus seems to have been reached that thermoelectric conversion can be viewed as reversible if considered independently of Joule heating and thermal conduction. [5,6,7,8]

Onsager's formulation of nonequilibrium thermodynamics[9] builds on the reciprocity of currents and flows of physical entities such as particles, entropy and heat which are considered stationary in classical thermodynamics. When applied to thermoelectricity, the irreversible features come to the fore, tied intimately with the resulting entropy generation.[10,11,12]

In this note we show, by extending Thomson's argument to a finite temperature difference, that thermoelectric conversion generating voltage in a substance with a non-zero Thomson coefficient implies irreversibility and entropy generation. The voltage is given by an analogue of Gouy-Stodola equation for classical heat engines.

$\pi(T)$

$T$

$w = q\delta V$

$T - \delta T$

$\delta Q$

$\pi(T - \delta T)$

Fig. 1. A schematic depiction of the thermoelectric generator operating between two temperatures $T$ and $T$ - $\delta T$.

## Thermoelectric converter as a heat engine

Following Thomson[2] we consider a thermoelectric converter operating between temperatures $T$ and $T$ - $\delta T$, where the temperature difference $\delta T$ small (Fig.1). If the current passing through the converter is negligible, we can perceive the thermoelectric as operating at open circuit. The converter can then be viewed as an energy (rather than power) converter - a classical heat engine which absorbs heat $\pi(T)$ per electron and carries out work $w$. In today's terminology, we can equate this work to the generated voltage $\delta V$ multiplied by the electron charge $q$.

If the engine/converter operates with the Carnot efficiency $\delta T/T$, we then have

$$w = q\delta V = \frac{\delta T}{T}\pi(T) \qquad (1)$$

Thomson realized that to balance the absorbed heat $\pi(T)$ with the work $w$ and the heat $\pi(T\text{-}\delta T)$ carried by an electron, a further heat $\delta Q$ has to be rejected into the reservoir at temperature $T$ - $\delta T$ :

$$\delta Q = \{\pi(T) - \pi(T-\delta T)\} - \frac{\delta T}{T}\pi(T) \simeq T\frac{d}{dT}\left(\frac{\pi}{T}\right)\delta T \qquad (2)$$

This is the Thomson heat.

With emphasis on reversibility, we consider the entropy balance by equating the ratio $\pi(T)/T$ to the entropy $s$ per electron. When the temperature is lowered from $T$ to $T$ - $\delta T$, the entropy of the thermoelectric per electron decreases by

$$\delta s' = \frac{\pi(T)}{T} - \frac{\pi(T-\delta T)}{T-\delta T} \simeq \frac{d}{dT}\left(\frac{\pi}{T}\right)\delta T \qquad (4)$$

At the same time, the rejection of heat $\delta Q$ into the low-temperature reservoir increases the entropy of the reservoir by

$$\delta s = \frac{\delta Q}{T-\delta T} \simeq \frac{d}{dT}\left(\frac{\pi}{T}\right)\delta T \qquad (3)$$

Since $\delta s = \delta s'$ (at least to first order in the temperature difference $\delta T$ ), there is no change in the total entropy of the system and the process can be considered reversible, consistent with our previous use of the Carnot efficiency to produce work.

By identifying $\pi$ as the Peltier heat and introducing the Seebeck coefficient $\varepsilon$ and Thomson coefficient $\tau$, we obtain

$$\varepsilon = \frac{q\delta V}{\delta T} = \frac{\pi}{T} = s \qquad (5)$$

from (1) and

$$\tau = \frac{\delta Q}{\delta T} = T\frac{d\varepsilon}{dT} = \frac{d\pi}{dT} - \varepsilon \qquad (6)$$

from (2) - in other words, Kelvin's two relations of thermoelectricity.

The thermoelectric conversion at small temperature difference $\delta T$ can be generalized to a finite difference $T_h$ - $T_c$ (Fig. 2). We note immediately that simply repeating the procedure outlined above requires many (effectively infinite number) of reservoirs at intermediate temperatures $T_h$ - $n\delta T$. But this contravenes the fundamental premise in the operation of classical heat engines where heat is absorbed from one reservoir at $T_h$, and rejected into a single reservoir at $T_c$.

Viewed in this way we find that the work is now

$$qV = \int_{T_c}^{T_h} \varepsilon \, dT \qquad (7)$$

and heat rejected into the reservoir at $T_c$ is

$$Q = \{\pi(T_h) - \pi(T_c)\} - \int_{T_c}^{T_h} \varepsilon \, dT = \int_{T_c}^{T_h} \tau \, dT \qquad (8)$$

where the second equation has been obtained by a simple integration by parts.

The entropy increase of the low-temperature reservoir is now $Q/T_c$ and entropy reduction of the thermoelectric $\varepsilon(T_h)$ - $\varepsilon(T_c)$, giving an overall entropy generation

$$S_i = \frac{Q}{T_c} - \{\varepsilon(T_h) - \varepsilon(T_c)\} = \frac{1}{T_c}\int_{T_c}^{T_h} \tau \, dT - \int_{T_c}^{T_h} \frac{d\varepsilon}{dT} dT = \frac{1}{T_c}\int_{T_c}^{T_h} \left(1 - \frac{T_c}{T}\right) \tau \, dT \qquad (9)$$

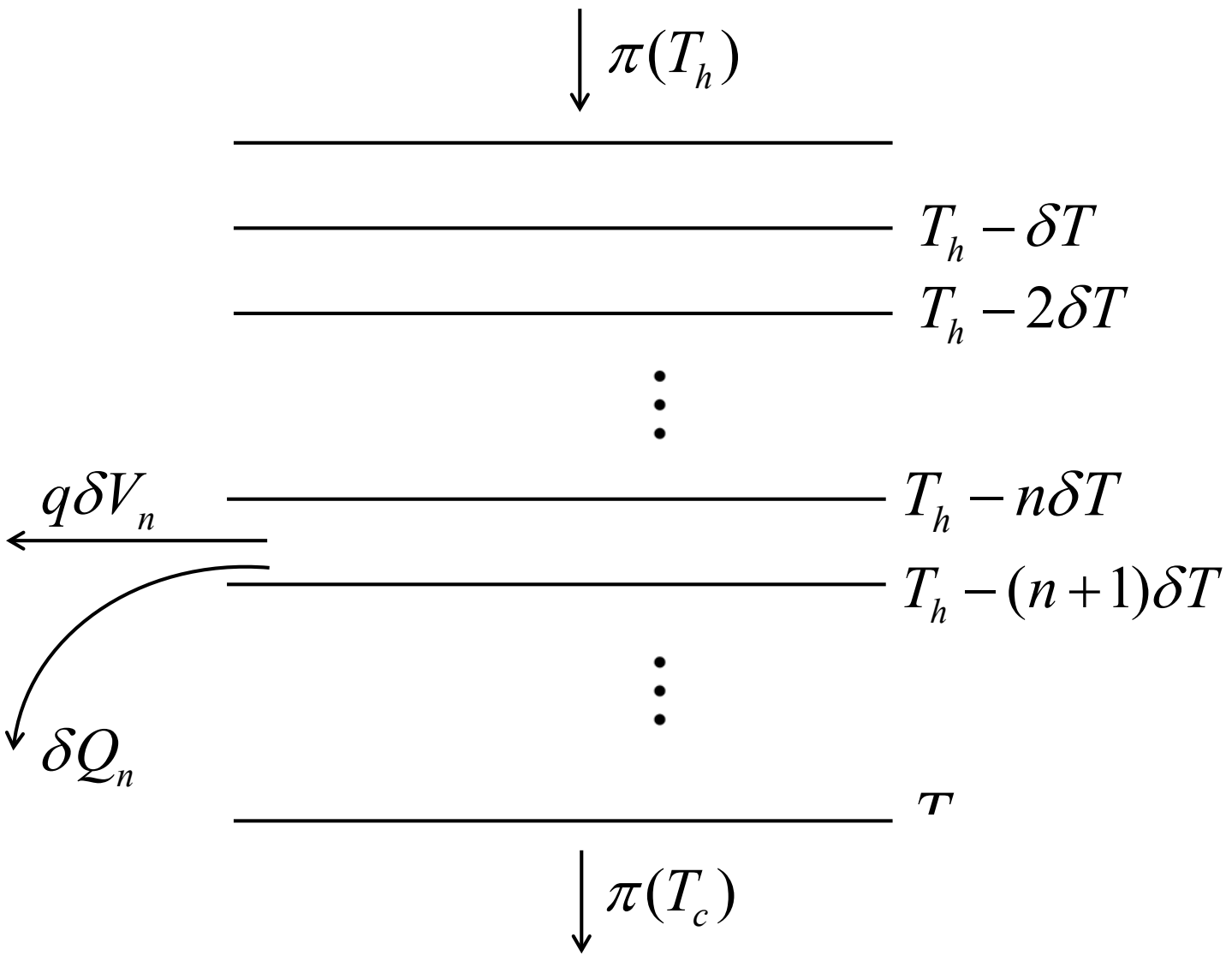

Fig.2 Thermoelectric generator operating between two temperatures $T_h$ and $T_c$.

Combining (8) and (9) and making use of Kelvin's second relation (5) we can now write down the reversible and irreversible contribution to the thermoelectric voltage:

$$qV = \{\pi(T_h) - \pi(T_c)\} - \{\varepsilon(T_h) - \varepsilon(T_c)\} T_c - T_c S_i = \left(1 - \frac{T_c}{T_h}\right)\pi(T_h) - T_c S \qquad (10)$$

In other words, the thermoelectric voltage is equal to the work carried out by the conversion of the heat absorbed from the high-temperature reservoir with the Carnot efficiency, minus the "lost work" due to the irreversible entropy generation (9). Equation (10) is just an expression of the Guy-Stodola theorem for classical heat engines, applied to the thermoelectric converter.

## Discussion and conclusion

By extending Kelvin's argument to a finite working temperature difference converted by the thermoelectric and employing entropy balance we have shown that, even at open circuit, the voltage generation by a thermoelectric generates entropy whenever the Thompson's coefficient is nonzero. The conversion process is then irreversible, and the Carnot efficiency cannot be reached even at open circuit, generalizing the previous result.[13]

The irreversible entropy generation (9) has been obtained by using the classical (equilibrium) thermodynamics, similar to electrochemical sources such as batteries or fuel cells where initial analysis is performed at open circuit. The losses are additional to the irreversibility losses that follow from Onsager's irreversible thermodynamics. Joule heat, for example, can be viewed as a loss resulting from finite-time operation of the classical (static) heat engine that we have considered at open circuit.

## Acknowledgements

This work was supported by project "Energy Conversion and Storage", grant no. CZ.02.01.01/00/22_008/0004617, programme Johannes Amos Commenius, Excellent Research.